\documentclass[aip,pop,groupedaddress,preprint]{revtex4-1}
\usepackage{physics}
\usepackage{amsmath}
\usepackage{graphicx}
\usepackage{subfigure}
\usepackage{hyperref}
\begin{document}
\title{The electron acoustic waves in plasmas with two kappa-distributed electrons at the same temperatures and immobile ions} 
\author{Ran Guo}
\thanks{Author to whom correspondence should be addressed}
\email{rguo@cauc.edu.cn}
\affiliation{Department of Physics, College of Science, Civil Aviation University of China, Tianjin 300300, China}
\pacs{}
\begin{abstract}
    The linear electron acoustic waves propagating in plasmas with two kappa-distributed electrons and stationary ions are investigated.
    The temperatures of the two electrons are assumed to be the same, but the kappa indices are not.
    It shows that if one kappa index is small enough and the other one large enough, a weak damping regime of the electron acoustic waves exists.
    The dispersions and damping rates are studied numerically.
    The parameter spaces for the weakly damped electron acoustic waves are analyzed.
    Besides, the electron acoustic waves in the present model are compared with those in other models, especially the plasmas with two-temperature electrons.
    At last, we perform the Vlasov-Poisson simulations to verify the theory.
\end{abstract}
\maketitle

\section{Introduction}
\label{sec:intro}
The electron acoustic wave (EAW) is a low-frequency mode compared with the Langmuir wave (LW) in electrostatic plasmas. 
It has been studied for many years and attracted lots of interest.
Fried and Gould firstly proposed the concept of EAWs and indicated this mode as a heavily damped solution in the Maxwellian plasmas. \cite{Fried1961} 
After that, a weak damping regime of the linear EAWs was found in the plasmas consisting of two-temperature Maxwellian electrons and immobile ions. \cite{Watanabe1977,Gary1985}
Valentini \textit{et al.} suggested that the nonlinear EAWs could be excited by an external resonant driver with a relatively low amplitude. \cite{Valentini2006}
In addition to the theoretical works, the EAWs have been observed experimentally in different laboratory plasmas. \cite{Montgomery2001,Anderegg2009,Chowdhury2017}

In space plasmas, the EAWs are essential for explaining the formation of the broadband electrostatic noises. \cite{Pottelette1999,Singh2001}
However, several recent observations indicate that the kappa distribution is more suitable than the Maxwellian one to model space plasmas, such as solar wind, \cite{Pierrard2016,Lazar2017a} solar corona, \cite{Vocks2008,Cranmer2014} and planetary magnetosphere. \cite{Schippers2008,Dialynas2009}
In the laboratory plasmas, Hellberg \textit{et al.} also confirmed that the kappa model can provide more suitable explanations of the observed EAW dispersion and damping rate than the bi-Maxwellian and Maxwellian-waterbag models. \cite{Hellberg2000}
The kappa velocity distribution is usually written as, \cite{Mace2009,Summers1991}
\begin{equation}
    f_\kappa(\vb{v})  \propto \left( 1 + \frac{v^2}{\kappa \theta^2} \right)^{-\kappa-1},
    \label{eq:kappa-dist}
\end{equation}
where $\theta$ is the most probable speed.
The temperature is defined in a kinetic manner,
\begin{equation}
    \frac{3}{2} k_B T = \int \frac{1}{2}mv^2 f_\kappa(\vb{v}) \dd{\vb{v}},
    \label{eq:def-T}
\end{equation}
where $T$ is the actual temperature in kappa-distributed plasmas. \cite{Hellberg2009,Livadiotis2009}
Such a temperature definition results in the relationship,
\begin{equation}
    \theta = \sqrt{\frac{\kappa-\frac{3}{2}}{\kappa} \frac{2k_BT}{m}}.
    \label{eq:theta}
\end{equation}
In the limit of $\kappa \rightarrow \infty$, the kappa distribution \eqref{eq:kappa-dist} recovers the Maxwellian distribution.
Therefore, the parameter $\kappa$ measures the distance away from the Maxwellian equilibrium.
The kappa distribution has been applied to describe various plasmas in numerous works, \cite{ElTaibany2017,Aravindakshan2018a,Wang2018,Scherer2020}
although its generation mechanism is still under discussion. \cite{Hasegawa1985,Vocks2003,Yoon2014,Livadiotis2019b,Guo2020,Guo2021}

The theory of EAWs has also been extended by adopting the kappa distribution.
Mace \textit{et al.} studied the EAWs in a plasma with hot kappa and cool Maxwellian electrons. \cite{Mace1999}
The authors found that the larger kappa index of hot electrons would produce the weaker damping in the acoustic regime.
Baluku \textit{et al.} developed the above work by considering both the hot and cool electrons are kappa-distributed but with different kappa values. \cite{Baluku2011}
They concluded that the damping rate mainly depends on the hot-to-cool temperature ratio, and the relative hot electron density determines the weak damping regime in wavenumbers. 
They also demonstrated that the effects of the kappa indices for hot and cool electrons are weak but nonnegligible. 
Besides, some other works related to the EAWs have also used the kappa distributions. \cite{Danehkar2011,Devanandhan2012,Rehman2017,Bansal2019} 

In this paper, we will show that the weakly damped EAWs can propagate in plasmas consisting of two kappa-distributed electrons and static ions.
The temperatures of the two-electron components can be the same, but the kappa indices must be different.
According to Eqs. \eqref{eq:kappa-dist} and \eqref{eq:theta}, the different kappa values and equal temperatures would lead to different most probable speeds for two electrons.
An adequate gap between the most probable speeds of two electrons would permit a weak damping regime of the EAWs.
It is a novel generation mechanism of the weakly damped EAWs due to the kappa distributions.

The paper is organized as follows.
In Sec. \ref{sec:model}, we introduce the new theoretical model and the corresponding dispersion equation.
In Sec. \ref{sec:num-analy}, we numerically solve the dispersion relation and damping rate of the EAWs.
The weak damping regime is analyzed according to the numerical dispersion solutions.
In Sec. \ref{sec:com}, our model of weakly damped EAWs is compared with some other ones, including the well-known two-temperature electron model and some other kappa-distributed models.
In Sec. \ref{sec:sim}, we conduct the Vlasov-Poisson simulations to verify the theory.
The results are summarized in the last Sec. \ref{sec:sum}.

\section{Model and Definition}
\label{sec:model}
We consider an electrostatic plasma consisting of two kappa-distributed electrons and immobile ions.
The ions are supposed to be spatially uniform to provide a neutral background.
The three-dimensional kappa distributions of the two electrons could be expressed as, \cite{Summers1991,Mace1995}
\begin{equation}
    f(\vb{v}) = \sum_{\sigma=s,f} \frac{n_\sigma}{(\kappa_\sigma\pi\theta_\sigma^2)^{3/2}} \frac{\Gamma(\kappa_\sigma+1)}{\Gamma(\kappa_\sigma-\frac{1}{2})} \left( 1+ \frac{v^2}{\kappa_\sigma \theta_\sigma^2} \right)^{-\kappa_\sigma-1},
    \label{eq:mul-kappa-dist}
\end{equation}
where, for species $\sigma$, $n_\sigma$ is the number density, $\kappa_\sigma$ is the kappa index, and $\theta_\sigma$ is the most probable speed.
The subscripts $\sigma = s,f$ denote two different electron components which we call slow and fast electrons, respectively.
We assume the temperature $T$ (defined by Eq. \eqref{eq:def-T}) of the two electrons is the same.
As a result, the two species have the different most probable speeds,
\begin{equation}
    \theta_\sigma = \sqrt{\frac{\kappa_\sigma-\frac{3}{2}}{\kappa_\sigma} \frac{2k_BT}{m}},
    \label{eq:mul-theta}
\end{equation}
due to the different kappa indices $\kappa_\sigma$.
We must stress that the range of the kappa index $\kappa_\sigma$ is required to be $(3/2,+\infty]$ to ensure the convergence of the second moment of the distribution \eqref{eq:mul-kappa-dist}. \cite{Livadiotis2010a}
It leads to that $\theta_\sigma$ must be restricted in the range $(0,\sqrt{2k_BT/m}]$. 
We suppose the kappa indices $\kappa_s<\kappa_f$ and thus the most probable speeds $\theta_s<\theta_f$. 
This is the reason for the name of slow and fast electrons.
For such electrostatic plasmas, the linear dispersion relation is, \cite{Summers1991,Mace1995}
\begin{equation}
     1 + \sum_{\sigma=s,f} \frac{2 \omega_\sigma^2}{k^2 \theta_\sigma^2} \left[ 1 - \frac{1}{2\kappa_\sigma} +\xi_\sigma Z(\kappa_\sigma;\xi_\sigma) \right]=0,
     \label{eq:pdr}
\end{equation}
where $\omega_\sigma = \sqrt{n_\sigma e^2/(m \varepsilon_0)}$ is the plasma frequency for each species, $k$ is the wavenumber, and $Z(\kappa_\sigma;\xi_\sigma)$ is the modified plasma dispersion function defined by, \cite{Mace1995}
\begin{equation}
    Z(\kappa_\sigma;\xi_\sigma) = \frac{\Gamma(\kappa_\sigma)}{\sqrt{\pi\kappa_\sigma}\Gamma\left(\kappa_\sigma-\frac{1}{2}\right)} 
    \int_{-\infty}^{+\infty} \frac{\left( 1+\frac{s^2}{\kappa_\sigma} \right)^{-\kappa_\sigma-1}}{s-\xi_\sigma} \dd{s}
    \label{eq:Z-int}
\end{equation}
with $\xi_\sigma = \omega/(k \theta_\sigma)$.
One can rewrite the above equation in the form of the hypergeometric function, \cite{Mace1995} i.e., 
\begin{align}
    Z(\kappa_\sigma;\xi_\sigma) = &i \frac{\left( \kappa_\sigma + \frac{1}{2} \right)\left( \kappa_\sigma - \frac{1}{2} \right)}{\kappa_\sigma^{3/2}(\kappa_\sigma+1)} \times \notag \\
    &{_2 F_1} \left[1,2\kappa_\sigma+2;\kappa_\sigma+2;\frac{1}{2}\left(1-\frac{\xi_\sigma}{i\sqrt{\kappa_\sigma}}  \right)\right].
    \label{eq:Z-2f1}
\end{align}
According to Eq. \eqref{eq:mul-theta}, a sufficiently small $\kappa_s$ and large $\kappa_f$ lead to a large enough gap between $\theta_s$ and $\theta_f$, providing the possibilities for the weakly damped EAWs with the phase speed $\theta_s<w/k<\theta_f$.
It is worth noting that the analytic solution of the EAW dispersion is difficult to derive in our model.
The reason is explained in Appendix \ref{sec:ap}.
Therefore, the solution of the EAW dispersion is obtained numerically throughout this paper.

\section{Numerical analysis}
\label{sec:num-analy}
In the following numerical analysis, we calculate the dispersion relation directly from Eq. \eqref{eq:Z-2f1}.
The total density of electrons is set as $n_0 = n_s + n_f = 1$.
We choose $n_s$ as an independent parameter and thus $n_f = 1 - n_s$.
Both the real wave frequency $\omega_r$ and the damping rate $\gamma$ are calculated in the unit of the total plasma frequency $\omega_{pe} = \sqrt{n_0 e^2/(m \varepsilon_0)}$.
The wavenumber is expressed in the dimensionless form of $k\lambda_s$, where $\lambda_\sigma = \sqrt{\varepsilon_0 k_B T/(n_\sigma e^2)}$ is the standard (Maxwellian) Debye length for species $\sigma$.
It is worth noting that the Debye length for kappa-distributed plasmas is given by, \cite{Livadiotis2014a,Livadiotis2019}
\begin{equation}
    \lambda_{\kappa\sigma} = \sqrt{\frac{\kappa_\sigma-3/2}{\kappa_\sigma-1/2}} \sqrt{\frac{\varepsilon_0 k_BT_\sigma}{n_\sigma e^2}} =\sqrt{\frac{\kappa_\sigma-3/2}{\kappa_\sigma-1/2}} \lambda_\sigma. 
    \label{eq:debye-len}
\end{equation}
However, we still use $k\lambda_{\sigma}$ as the dimensionless wavenumber in this study.
The reason is that we want to compare the dispersions and damping rates of different $\kappa_\sigma$ for the same wavenumbers.
If we adopt $k\lambda_{\kappa \sigma}$ as the dimensionless wavenumber, 
then for the same $k\lambda_{\kappa \sigma}$, the wavenumber $k$ is different when the different $\kappa_\sigma$ is taken.
The comparisons of the dispersions and damping rates for different wavenumbers $k$ are meaningless.
\begin{figure}[ht]
	\centering
    \includegraphics[width=\textwidth]{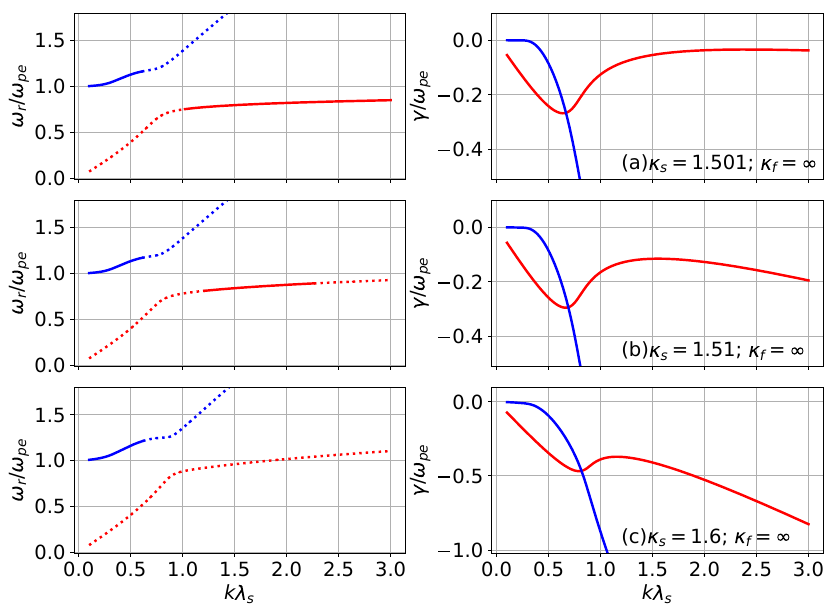}
    \caption{
        The dispersion relation (left panel) and damping rate (right panel) of the LWs and EAWs for varied $\kappa_s$.
        The LWs are denoted by blue, while the EAWs by red.
        The solid lines represent the weak damping regime, namely $|\gamma|<\omega_r/(2\pi)$. 
        The strong damping regime, identified by $ |\gamma|\ge \omega_r/(2\pi)$, is drawn by the dotted lines.
        We set $\kappa_f = \infty$ to avoid the strongly damped EAWs caused by a small $\kappa_f$.
        The influence of $\kappa_s$ is shown by varying its value from $1.501$ to $1.6$.
        The number density of slow electrons is selected as $n_s = 0.7$.
    }
    \label{fig:fig1}
\end{figure}
\begin{figure}[ht]
	\centering
    \includegraphics[width=\textwidth]{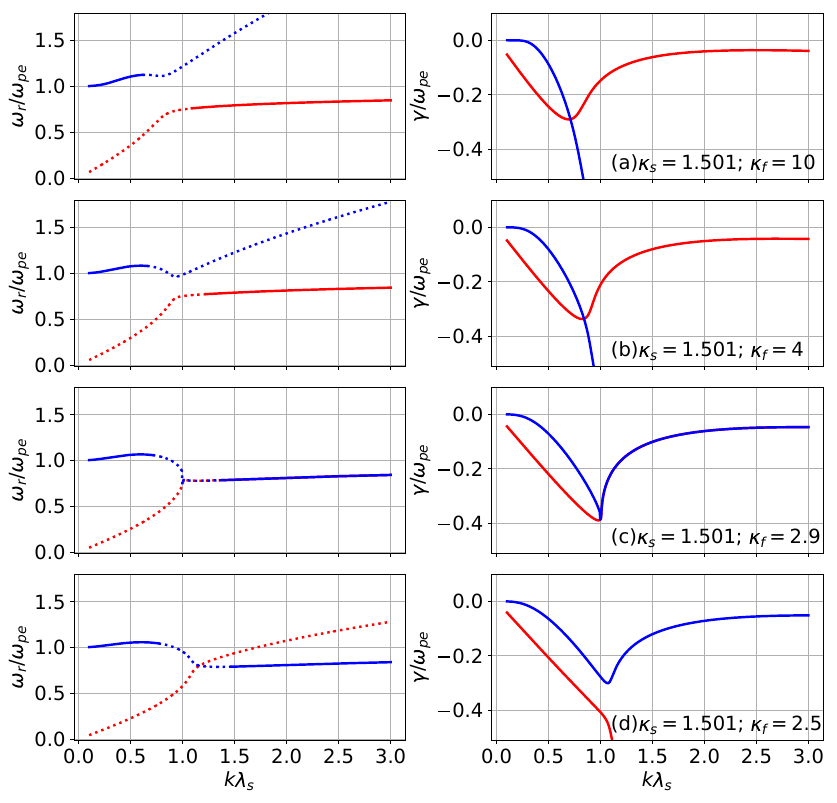}
    \caption{
        The dispersion relation (left panel) and damping rate (right panel) of the LWs and EAWs for varied $\kappa_f$.
        The legends are the same as those in Fig. \ref{fig:fig1}.
        We set $\kappa_s = 1.501$ to avoid the strongly damped EAWs due to a large $\kappa_s$.
        The influence of $\kappa_f$ is shown by varying its value from $2.5$ to $10$.
        The number density of slow electrons is still selected as $n_s = 0.7$.
        }
    \label{fig:fig2}
\end{figure}

Figs. \ref{fig:fig1} and \ref{fig:fig2} are drawn to illustrate the existence of the weakly damped EAWs due to the different kappa indices.
In Fig. \ref{fig:fig1}, one finds that the EAWs would have a weak damping regime if $\kappa_s$ is small enough.
The weak damping range in wavenumbers expands when $\kappa_s$ tends to its minimum $3/2$.
Fig. \ref{fig:fig2} indicates that a large enough $\kappa_f$ allows the EAWs with weak damping.
During the reduction of $\kappa_f$, the dispersion curves of LW and EAW are close to each other.
In the left panel of Fig. \ref{fig:fig2}(c), when $\kappa_f$ approaches the critical value $2.9$, the two dispersion curves overlap for approximately $k\lambda_s>1.0$.
The damping curves have a similar situation in the right panel of Fig. \ref{fig:fig2}(c).
For $\kappa_f<2.9$, the LW and EAW branches cross, as shown in the left panel of Fig. \ref{fig:fig2}(d).
The wavenumber of the intersection point is $k\lambda_s \approx 1.15$.
For $k\lambda_s > 1.15$, the upper branch of the dispersion is identified as the EAW, while the lower branch is the LW.
The reason is that the damping rate must be continuous for each branch.
Besides, one can also find that the weak damping regime in wavenumbers does not change obviously for a decreasing $\kappa_f$ before the two branches cross.

\begin{figure}[ht]
	\centering
    \includegraphics[width=0.5\textwidth]{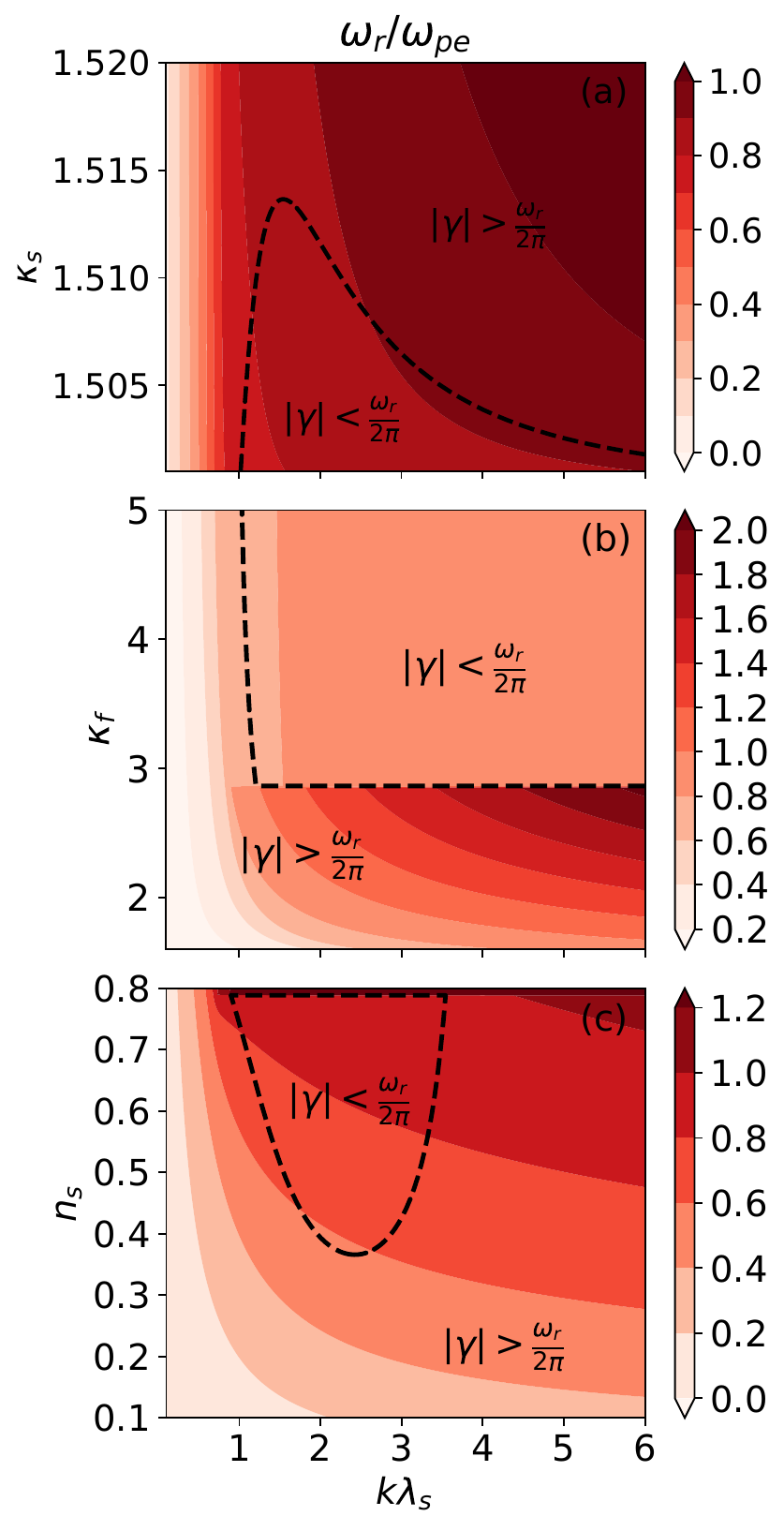}
    \caption{
        The EAW dispersion and the weak damping regime for variant (a) $\kappa_s$, (b) $\kappa_f$, and (c) $n_s$.
        The color bar indicates the value of $\omega_r/\omega_{pe}$ in all three subfigures.
        The dashed lines are the separatrices of the strong and weak damping regimes.
        The other parameters used in the subplots are (a) $n_s = 0.7$ and $\kappa_f = \infty$, 
        (b) $n_s = 0.7$ and $\kappa_s = 1.501$, 
        and (c) $\kappa_s = 1.505$ and $\kappa_f = \infty$. 
        }
    \label{fig:fig3}
\end{figure}

Fig. \ref{fig:fig3} exhibits the dispersion and accurate weak damping regime of the EAWs.
In Fig. \ref{fig:fig3}(a), we find that if $\kappa_s$ varies with fixed $n_s=0.7$ and $\kappa_f=\infty$, the real wave frequency $\omega_r$ changes in large wavenumbers (roughly $k\lambda_s > 1$) but is nearly unchanged in small wavenumbers (roughly $k\lambda_s < 1$).
The range of $\kappa_s$ for weakly damped EAWs is about $3/2<\kappa_s<1.514$ in this case.
It implies $\kappa_s$ must be very close to its minimum $3/2$; otherwise, there is no weak damping regime.
From Fig. \ref{fig:fig3}(b), we find that $\kappa_f$ has a lower limit $\kappa_f \approx 2.9$ in the weak damping regime. 
For $\kappa_f > 2.9$, both the dispersion and the wavenumbers of the weak damping regime remain almost the same.
It means that $\kappa_f$ is less important if it is larger than some critical value.
Such a discontinuity at $\kappa_f \approx 2.9$ could be attributed to the cross of the LW and EAW branches.
When $\kappa_f > 2.9$, just like the case shown in Fig. \ref{fig:fig2}(b), the two branches do not cross, and there is a weak damping regime for EAWs.
But, when $\kappa_f<2.9$, as shown in Fig. \ref{fig:fig2}(d), the two branches cross.
In this case, the weak damping regime belongs to the LW rather than the EAW due to the continuity of the damping curve.
Therefore, there is no weakly damped EAW for $\kappa_f<2.9$.
In Fig. \ref{fig:fig3}(c), the weakly damped EAWs can only exist for approximately $0.36<n_s<0.79$ with $\kappa_s = 1.505$ and $\kappa_f=\infty$.
The weak damping regime in wavenumbers is also affected by different $n_s$.

\begin{figure}[ht]
	\centering
    \includegraphics[width=0.5\textwidth]{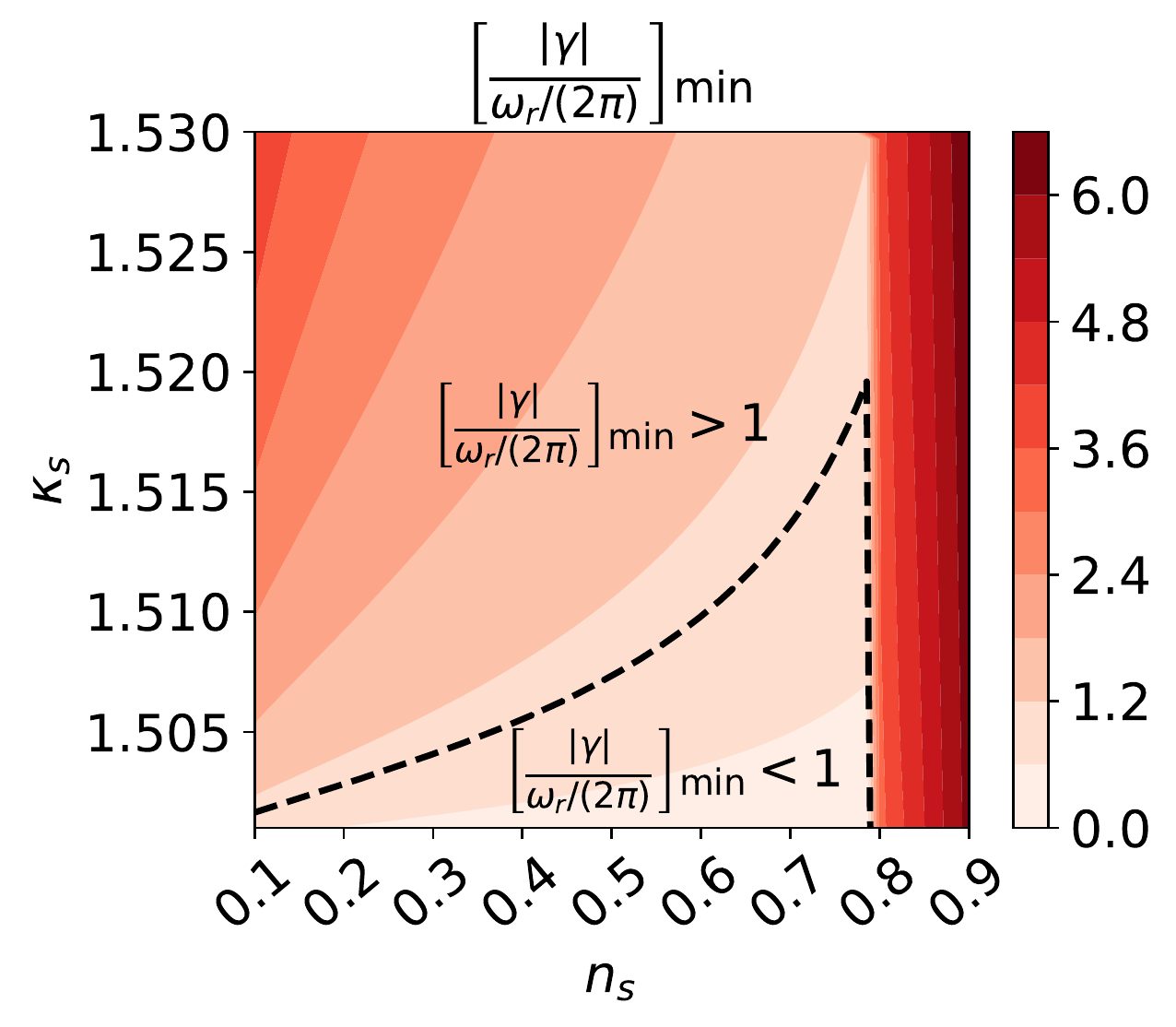}
    \caption{
        The minimum of $\frac{|\gamma|}{\omega_r/(2\pi)}$ and the weak damping regime. 
        The color bar indicates the minimum value of $\frac{|\gamma|}{\omega_r/(2\pi)}$ in the range of $0.0<k\lambda_s<6.0$.
        We let $\kappa_f = \infty$ in this figure.
        The dashed line is the separatrix that $\left[\frac{|\gamma|}{\omega_r/(2\pi)}\right]_{\min} = 1$.
        }
    \label{fig:fig4}
\end{figure}

In terms of the above analysis, it is clear that $\kappa_s$ and $n_s$ are the main factors affecting the EAW damping.
Therefore, we investigate the weak damping regime of the EAWs in the parameter spaces of $\kappa_s$ and $n_s$.
We use the minimum of $2\pi|\gamma|/\omega_r$ in a certain wavenumber range as the criterion for the existence of weakly damped EAWs.
If $[2\pi|\gamma|/\omega_r]_{min}<1$, there must be a weak damping regime in the corresponding wavenumbers.
The results are illustrated in Fig. \ref{fig:fig4}.
It shows a larger $n_s$ (in the range of $0.1<n_s<0.79$ approximately) enlarges the range of $\kappa_s$ for the weakly damped EAWs.
Nevertheless, the range of $\kappa_s$ is still very small.
The maximum of $\kappa_s$ is about $1.519$ at $n_s \approx 0.79$.
Some observations and analyses indicate that the kappa index could be very close to its minimum $3/2$. \cite{STEFFL2004,Decker2005,Fahr2015}
In Ref. \onlinecite{STEFFL2004}, Steffl \textit{et al.} derived the kappa value from the observations of Io plasma torus by assuming the kappa-distributed electrons.
They found that the kappa value could decrease to $\kappa \approx 1.5$ at a long radial distance.
Decker \textit{et al.} \cite{Decker2005} analyzed the observation data of the heliospheric termination shock.
Their work suggested that the protons could be described by the kappa distribution with $\kappa = 1.63$.
Fahr \textit{et al.} \cite{Fahr2015} studied the electron distribution downstream of the solar wind termination shock and found the electrons should follow the kappa distribution with $\kappa = 1.522$.

\section{Comparison with the other models}
\label{sec:com}
\subsection{Comparison with the model of the two-temperature electrons}
\begin{figure}[ht]
	\centering
    \includegraphics[width=0.5\textwidth]{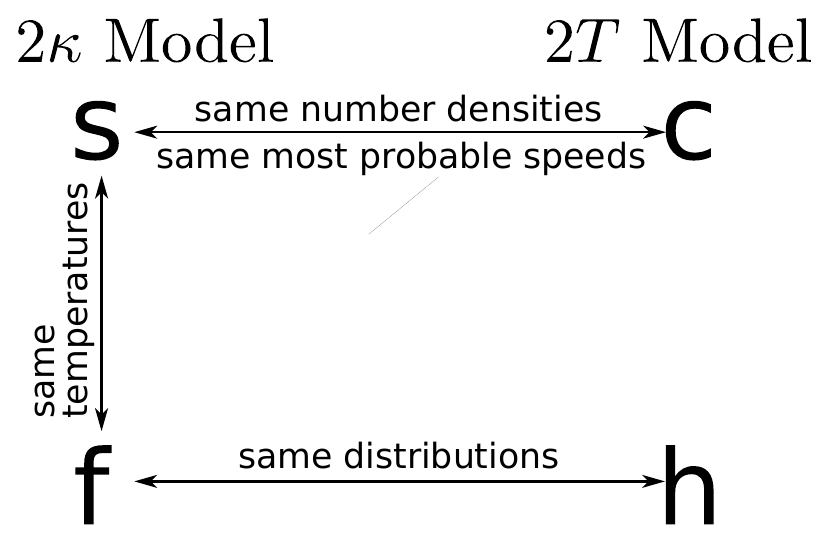}
    \caption{
        Schematic diagram of the relationships between the two-kappa-distribution model ($2\kappa$ Model) and the two-temperature model ($2T$ Model) in our comparison.
        The notations 's', 'f', 'c', and 'h' represent the slow, fast, cold, and hot electrons, respectively, in these two models.
        }
    \label{fig:fig5}
\end{figure}

As we know, the EAWs are weakly damped in the plasmas with two-temperature electrons and stationary ions. \cite{Watanabe1977,Gary1985}
From the analysis in Sec. \ref{sec:num-analy}, we can infer that the slow electrons and fast electrons in our model play the similar roles as the cold and hot electrons in the two-temperature model, respectively.
Therefore, it is necessary to compare the EAWs in these two different models.
To distinguish them, the model proposed in this paper is called the two-kappa-distribution model, and the model in the previous references \cite{Watanabe1977,Gary1985} is called the two-temperature model.
The one-dimensional distribution for the two-temperature model is,
\begin{equation}
    f_{2T}(v) = \sum_{\sigma=c,h} \frac{n_\sigma}{\sqrt{\pi\theta_\sigma^2}} e^{-\frac{v^2}{\theta_\sigma^2}},
    \label{eq:pdf-2T}
\end{equation}
where the subscripts $c, h$ denote the cold and hot electrons, respectively.
By integrating Eq. \eqref{eq:mul-kappa-dist}, one derives the one-dimensional distribution for the two-kappa-distribution model,
\begin{equation}
    f_{2\kappa}(v) = \sum_{\sigma=s,f} \frac{n_\sigma}{\sqrt{\kappa_\sigma\pi\theta_\sigma^2}}\frac{\Gamma(\kappa_\sigma)}{\Gamma\left(\kappa_\sigma-\frac{1}{2}\right)} \left( 1+\frac{v^2}{\kappa_\sigma \theta_\sigma^2} \right)^{-\kappa_\sigma}.
    \label{eq:pdf-2kappa}
\end{equation}
For comparison, we set the most probable speeds and the number densities are the same for the two models, i.e., $\theta_c=\theta_s$, $\theta_h = \theta_f$, $n_c=n_s$, and $n_h=n_f$.
In addition, according to the analysis in Sec. \ref{sec:num-analy},  $\kappa_f$ does not play an important role, so we let $\kappa_f = \infty$.
It means the fast electron distribution is the same as the hot electron distribution in this comparison.
We treat $\theta_s$ and $\theta_f$ as independent parameters, so $\kappa_s$ is determined by,
\begin{equation}
    \kappa_s = \frac{3}{2\left(1-\frac{\theta_s^2}{\theta_f^2}\right)},
    \label{eq:com-kappas}
\end{equation}
to ensure the slow and fast electrons have the same temperatures.
Under these constraints, the only difference between the two models is the distribution functions for cold electrons and slow electrons.
The relationships between the two models are drawn in Fig. \ref{fig:fig5}.
\begin{figure}[ht]
	\centering
    \includegraphics[width=0.5\textwidth]{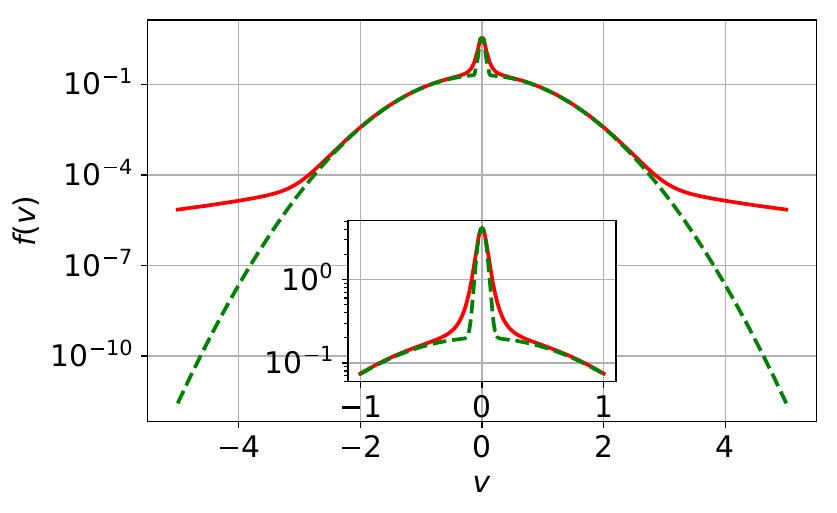}
    \caption{
        The distributions for the two-temperature and the two-kappa-distribution models.
        The green dashed lines stand for the two-temperature distribution \eqref{eq:pdf-2T} and the red solid line for the two-kappa distribution \eqref{eq:pdf-2kappa}.
        We set the number densities as $n_c = n_s = n_h = n_f = 0.5$ and the most probable speeds as $\theta_c = \theta_s = 0.05$, $\theta_h = \theta_f = 1$.
        The kappa index $\kappa_s \approx 1.504$ is calculated from Eq. \eqref{eq:com-kappas}.
        }
    \label{fig:fig6}
\end{figure}
\begin{figure}[ht]
	\centering
    \includegraphics[width=0.5\textwidth]{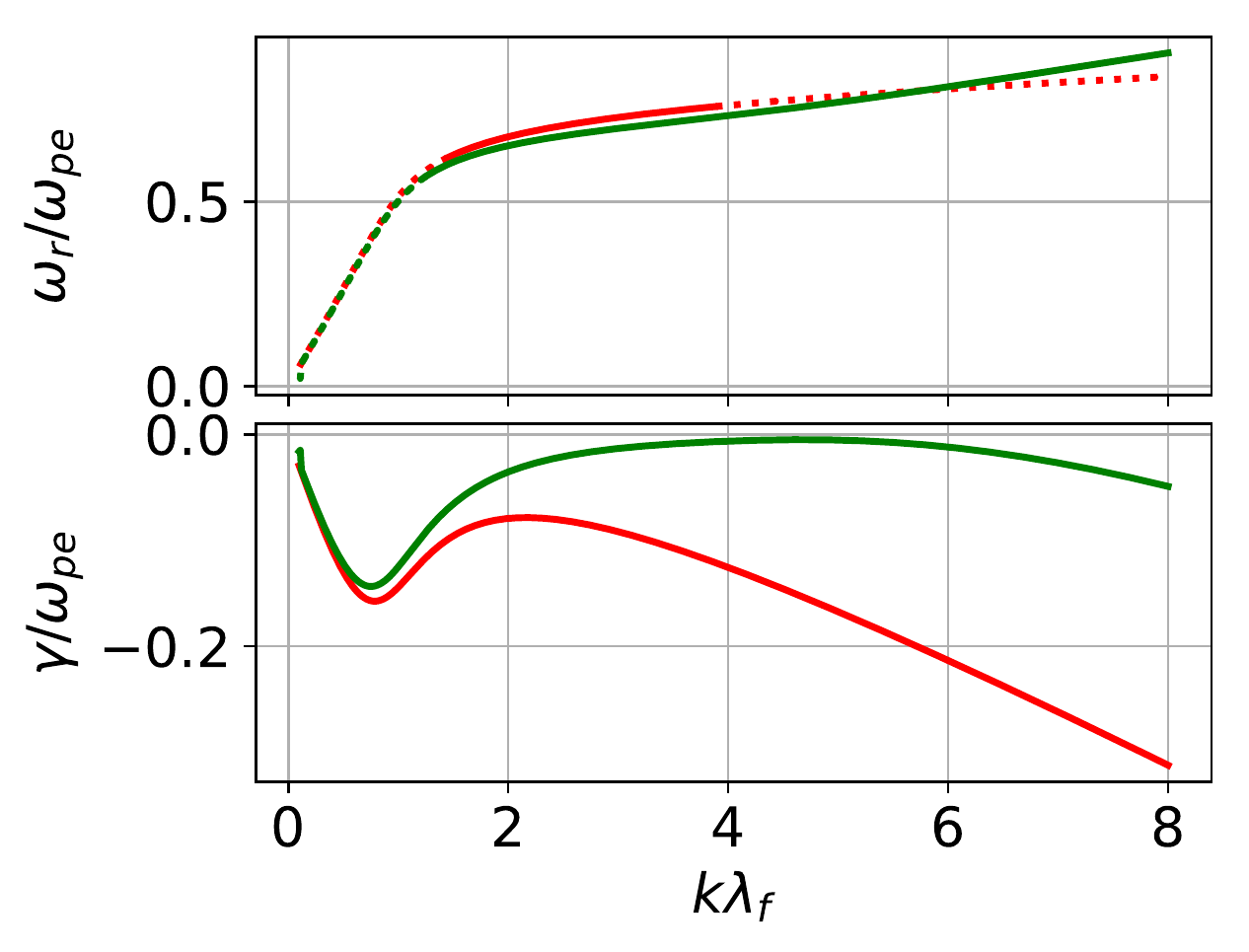}
    \caption{
        The dispersion and damping rate of the EAWs in the two-temperature model (green lines) and the two-kappa-distribution model (red lines).
        The dotted lines in the upper panel denote the strong damping regime, while the solid lines denote the weak one.
        The parameters used in this figure are the same as those in Fig. \ref{fig:fig6}.
        }
    \label{fig:fig7}
\end{figure}

For comparison, we plot the distributions for these two models in Fig. \ref{fig:fig6}.
It shows that the two-kappa-distribution model has more suprathermal electrons in high energies.
In low energies, the two-kappa distribution is a little fatter than the two-temperature one.
The dispersion and damping rate are illustrated in Fig. \ref{fig:fig7}.
We find that the two dispersion curves are very close to each other, but the weak damping regime of the two-kappa-distribution model is smaller than that of the two-temperature model in wavenumbers.
The reason is that the damping strength of the two-kappa-distribution model is stronger than the two-temperature model, which is displayed in the lower panel of Fig. \ref{fig:fig7}.

\subsection{Comparison with the other kappa-distributed models}
There are some other works \cite{Mace1999,Baluku2011} studying the EAWs in kappa-distributed plasmas.
However, the present study is different from them.
In Ref. \onlinecite{Mace1999}, the authors investigated the EAWs in the plasmas with hot kappa and cool Maxwellian electrons.
In their model, the kappa-distributed electrons are assumed to be much hotter than the Maxwellian ones.
Because the kappa index in their paper is not very small (the minimum kappa value may be $1.6$), we can infer that the most probable speed of the hot kappa electron is still much larger than that of the cool Maxwellian electron.
Therefore, in their work, the slow species (cool Maxwellian electrons) has a larger kappa index ($\kappa \rightarrow \infty$) than the fast species (kappa-distributed electrons with a finite kappa index), which is very different from the present study.
In Ref. \onlinecite{Baluku2011}, Baluku \textit{et al.} studied the EAWs propagating in the plasmas with hot and cool kappa-distributed electrons.
In their work, they assumed the $\lambda_c \ll \lambda_{\kappa h}$, where $\lambda_c$ is the standard Debye length for the cool electrons while $\lambda_{\kappa h}$ is the Debye length defined by Eq. \eqref{eq:debye-len} for the hot species.
However, such an assumption cannot be valid if the two species have equivalent temperatures.
Therefore, the results obtained in Ref. \onlinecite{Baluku2011} are not appropriate for our model.

\section{Verification by Vlasov-Poisson simulations}
\label{sec:sim}
To verify the weakly damped EAWs in the two-kappa-distribution model, we perform the one-dimensional Vlasov-Poisson simulation. \cite{Cheng1976}
The time evolution of electrostatic plasmas follows the Vlasov-Poisson equations,
\begin{equation}
    \pdv{f}{t} + v\pdv{f}{x} -\frac{eE}{m}\pdv{f}{v} =0,
    \label{eq:V}
\end{equation}
\begin{equation}
    \pdv[]{E}{x} = \frac{e}{\varepsilon_0}\left( n_0 - \int_{-\infty}^\infty f \dd{v} \right),
    \label{eq:P}
\end{equation}
where $E$ is the electric field. The ion number density is set as $n_0$, providing a neutralizing background.
We introduce an initial perturbation by,
\begin{equation}
    f(x,v,t=0) = [1+d\cos(kx)] f_{2\kappa}
\end{equation}
with the small disturbance $d=0.001$ and the two-kappa distribution $f_{2\kappa}$ given by Eq. \eqref{eq:pdf-2kappa}.
In the simulation, the kappa indices are chosen as $\kappa_s=1.501$ and $\kappa_f=20$.
The number densities for the slow and fast electrons are $n_s=0.75$ and $n_f=0.25$.
Besides, for both two species, the temperature is $T=1$, the electron mass $m=1$, the elementary charge $e=1$, and the vacuum permittivity $\varepsilon_0=1$.
Thus, the total plasma frequency is $\omega_{pe} = \sqrt{n_0 e^2/(m\varepsilon_0)} = 1$.

In the simulations, the Vlasov equation \eqref{eq:V} is integrated by the semi-Lagrangian splitting scheme with cubic spline interpolations. \cite{Cheng1976}
The Poisson equation \eqref{eq:P} is solved by the tridiagonal matrix algorithm. \cite{Press2007}
The simulation scale is $[0, L]$ with $L=2\pi/k$, and the periodic boundary conditions are adopted in the position space.
The simulation domain for the velocity space is $[-v_{max},v_{max}]$ with $v_{max}=10$, and outside this interval ($|v|>v_{max}$), the distribution is treated as zero.
The position space is discretized by $N_x = 200$ grid points and the velocity space by $N_v=4000$ grid points.
We can calculate the nonlinear trapping time approximately by $\tau = 2\pi\sqrt{m/(ed)} \approx 198.7$, \cite{Oneil1965}
so the simulation time is limited to $[0,t_{max}]$ with $t_{max}=40<\tau$ to ensure the linear evolution of the system. 
We select the time step as $\Delta t=0.005$.

\begin{figure}[ht]
	\centering
    \includegraphics[width=0.5\textwidth]{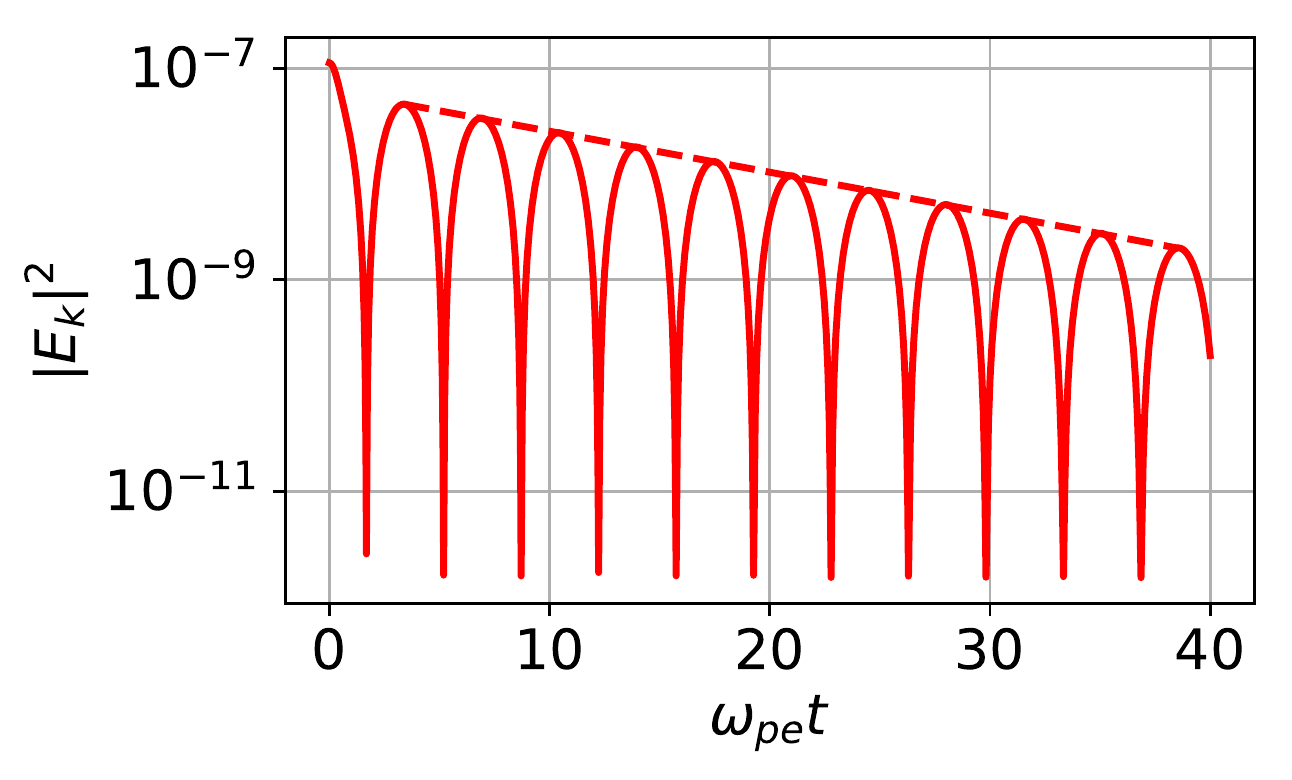}
    \caption{
        The time evolution of the amplitude of the electric field $E_k$ for the wavenumber $k\lambda_s=3.5$.
        The solid lines are the simulation results, while the dashed lines are the numerical solutions from the dispersion \eqref{eq:Z-2f1}.
        }
    \label{fig:fig8}
\end{figure}
\begin{figure}[ht]
	\centering
    \includegraphics[width=0.5\textwidth]{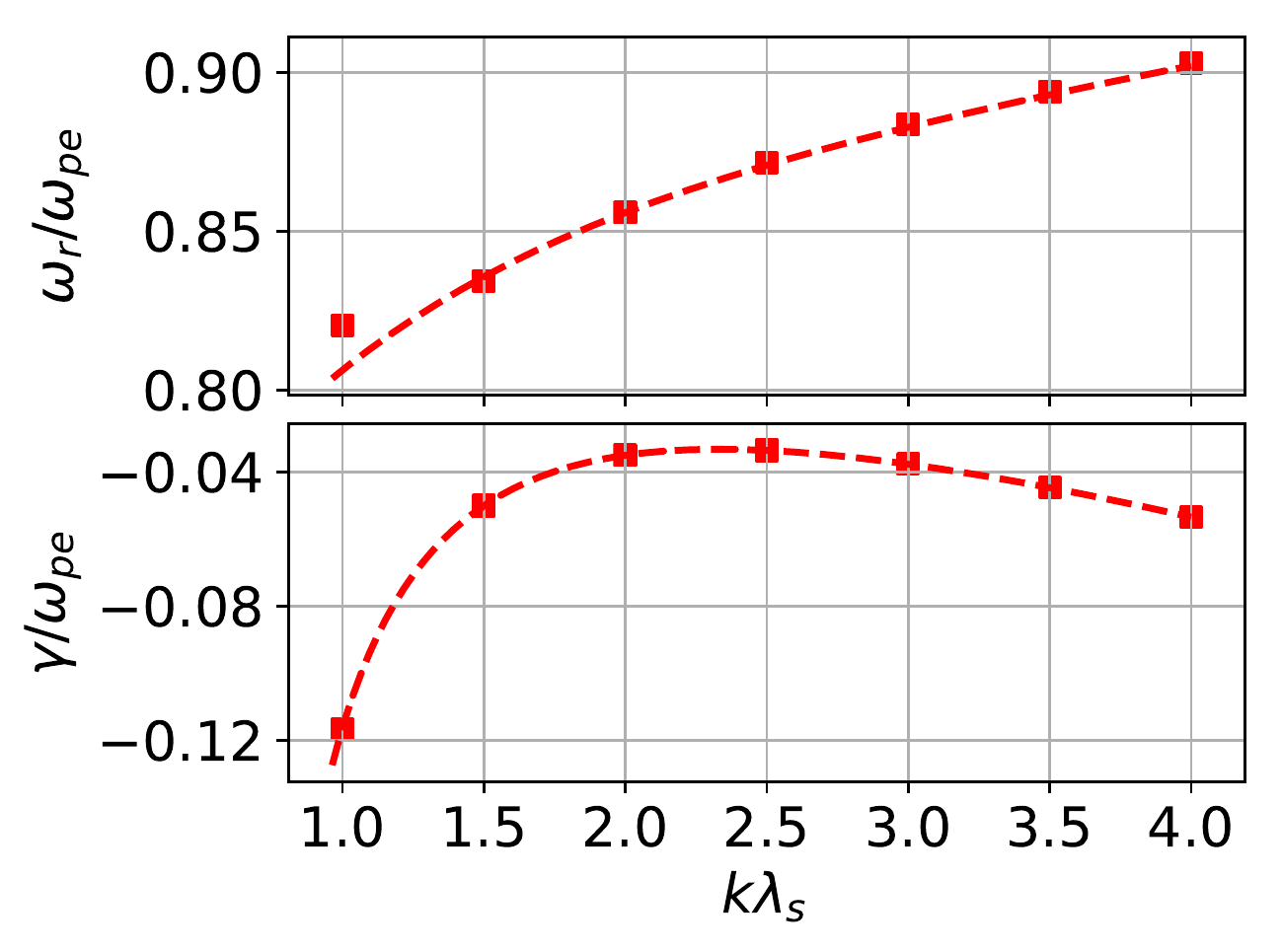}
    \caption{
        Verification of the wave frequency and damping rate in the weak damping regime $1.0<k\lambda_s<4.0$.
        The dashed lines stand for the numerical solutions from Eq. \eqref{eq:Z-2f1}, while the squares denote the simulation results.
        }
    \label{fig:fig9}
\end{figure}

The results are illustrated in Figs. \ref{fig:fig8} and \ref{fig:fig9}.
In Fig. \ref{fig:fig8}, we plot the time evolution of the electric field in the case of $k \lambda_s=3.5$.
One can obtain the wave frequency $\omega_r^{sim}=0.893767 $ and the damping rate $\gamma^{sim}=-0.044518 $ from the simulations.
As a comparison, the numerical calculations give $\omega_r^{num}=0.892928$ and $\gamma^{num}=-0.044569$ in terms of Eq. \eqref{eq:Z-2f1}.
The relative errors between the simulations and numerical results are about $0.094\%$ and $0.113\%$ for the wave frequency and damping rate, respectively.
In a large range of the wavenumbers, we illustrate the comparison of the wave frequency and damping rate in Fig. \ref{fig:fig9}. 
One finds that the simulations are in very good agreement with the numerical results.

\section{Summary}
\label{sec:sum}
In this paper, we investigate the weakly damped EAWs in plasmas consisting of two kappa-distributed electrons and stationary ions.
We assume the two-electron components have the same temperatures but the different kappa indices.
It is found that the EAWs in such plasmas would be weakly damped for sufficiently small $\kappa_s$ and large $\kappa_f$.
The kappa indices and the number densities for both two electrons are the dominant factors determining the weak damping regime in wavenumbers.
The parameter spaces permitting the weakly damped EAWs are illustrated in Figs. \ref{fig:fig3} and \ref{fig:fig4}.
In addition, we mainly compare the EAWs in our model and the well-known two-temperature electron model.
As exhibited in Fig. \ref{fig:fig7}, the different models would affect the dispersion curve slightly but the damping rate strongly.
It shows that the EAWs in our model have a smaller weak damping regime.  
At last, the Vlasov-Poisson simulations are performed for verifications.
Simulation results are in good agreement with the theory.

Our results imply that the most probable speed is the main factor determining the possibility of the weakly damped EAWs. 
For the plasmas with two Maxwellian electrons, the most probable speed is only related to the temperature.
Therefore, the temperature becomes the decisive factor in the Maxwellian plasmas.
However, in the plasmas with two kappa-distributed electrons, the most probable speed is not only related to the temperature but also the kappa index.
Thus, the temperature difference is not a necessary condition in the kappa-distributed plasmas.

\begin{acknowledgments}
This work was supported by the Fundamental Research Funds for the Central Universities, Civil Aviation University of China (No.3122019138).
\end{acknowledgments}

\section*{Data Availability}
The data that support the findings of this study are available from the corresponding author upon reasonable request.

\appendix
\section{Difficulties in the derivations of the analytic dispersion}
\label{sec:ap}
In this section, we demonstrate that even in the case of $\theta_s \ll w_r/k \ll \theta_f$,
the EAW dispersion cannot be obtained analytically.
The standard procedure for deriving the analytic EAW dispersion is to expand the modified dispersion functions $Z(\kappa_s;\xi_s)$ for $\xi_s \gg 1$ and $Z(\kappa_f;\xi_f)$ for $\xi_f \ll 1$.
Due to the weak damping, we neglect the image part of $\xi_\sigma$ and only consider the real part of the dispersion,
\begin{equation}
    1 + \sum_{\sigma=s,f} \frac{2 \omega_\sigma^2}{k^2 \theta_\sigma^2} 
    \left\{ 1 - \frac{1}{2\kappa_\sigma} +\xi_\sigma \Re[Z(\kappa_\sigma;\xi_\sigma)] \right\}
    =0.
    \label{eq:pdr-re}
\end{equation}
Let us consider the expansion of $\Re[Z(\kappa_s,\xi_s)]$ for $\xi_s \gg 1$.
According to Ref. \onlinecite{Mace1995}, the real part of $Z(\kappa_s;\xi_s)$ can be rewritten as, 
\begin{equation}
    \Re[Z(\kappa_s,\xi_s)] = - \xi_s \frac{( 2\kappa_s+1) (2\kappa_s-1)}{2\kappa_s^2} 
    {}_2 F_1\left( 1,\kappa_s+\frac{3}{2};\frac{3}{2};-\frac{\xi_s^2}{\kappa_s} \right).
    \label{ap:eq-Zr}
\end{equation}
In terms of linear transformations of the hypergeometric function, \cite{Olver2010}
\begin{align}
    \frac{\sin[\pi(b-a)]}{\pi} \frac{{}_2F_1(a,b;c;z)}{\Gamma(c)}
    =& \frac{(-z)^a {}_2F_1\left( a,a-c+1;a-b+1;\frac{1}{z} \right)}{\Gamma(b)\Gamma(c-a)\Gamma(a-b+1)} \notag \\
    &-\frac{(-z)^b {}_2F_1\left( b,b-c+1;b-a+1;\frac{1}{z} \right)}{\Gamma(a)\Gamma(c-b)\Gamma(b-a+1)},
\end{align}
for $|\arg(-z)|<\pi$, we have,
\begin{align}
     \sin\left( \kappa_s \pi + \frac{\pi}{2} \right) {}_2 F_1\left( 1,\kappa_s+\frac{3}{2};\frac{3}{2};-\frac{\xi_s^2}{\kappa_s} \right)
    =& \frac{\pi^{3/2}}{2\Gamma\left( \kappa_s+\frac{3}{2} \right)} 
     \left[ 
        \frac{\kappa_s}{\xi_s^2} \frac{{}_2F_1\left( 1,\frac{1}{2};\frac{1}{2}-\kappa_s;-\frac{\kappa_s}{\xi_s^2} \right)}{\sqrt{\pi}\Gamma\left( \frac{1}{2}-\kappa_s \right)}
     \right. \notag \\
     &\left.
        -\left( \frac{\kappa_s}{\xi_s^2} \right)^{\kappa_s+\frac{3}{2}} \frac{{}_2F_1\left(\kappa_s+\frac{3}{2},\kappa_s+1;\kappa_s+\frac{3}{2};-\frac{\kappa_s}{\xi_s^2} \right)}{\Gamma(-\kappa_s)}
     \right], 
     \label{ap:eq-ex-dev}
\end{align}
for $|\arg(\xi_s^2/\kappa_s)|<\pi$.
For non-half-integer $\kappa_s$, which means $\sin(\kappa_s \pi + \pi/2) \neq 0$, we can divide Eq. \eqref{ap:eq-ex-dev} by $\sin(\kappa_s \pi + \pi/2)$ on both sides, 
\begin{align}
     {}_2 F_1\left( 1,\kappa_s+\frac{3}{2};\frac{3}{2};-\frac{\xi_s^2}{\kappa_s} \right) 
    =& \frac{\pi^{3/2}}{2\cos(\kappa_s\pi)\Gamma\left( \kappa_s+\frac{3}{2} \right)} 
     \left[ 
        \frac{\kappa_s}{\xi_s^2} \frac{{}_2F_1\left( 1,\frac{1}{2};\frac{1}{2}-\kappa_s;-\frac{\kappa_s}{\xi_s^2} \right)}{\sqrt{\pi}\Gamma\left( \frac{1}{2}-\kappa_s \right)}
     \right. \notag \\
     &\left.
        -\left( \frac{\kappa_s}{\xi_s^2} \right)^{\kappa_s+\frac{3}{2}} \frac{{}_2F_1\left(\kappa_s+\frac{3}{2},\kappa_s+1;\kappa_s+\frac{3}{2};-\frac{\kappa_s}{\xi_s^2} \right)}{\Gamma(-\kappa_s)}
     \right]. 
     \label{ap:eq-ex-dev2}
\end{align}
Therefore, the above equation holds for non-half-integer $\kappa_s$ and otherwise diverges due to $\cos(\kappa_s\pi)$ on the denominator.
Substituting Eq. \eqref{ap:eq-ex-dev2} into \eqref{ap:eq-Zr} and expanding the hypergeometric functions, one derives the expansion of $\Re[Z(\kappa_s;\xi_s)]$ for non-half-integer $\kappa_s$,
\begin{align}
    \Re[Z(\kappa_s,\xi_s)] =&-\frac{\sqrt{\pi}}{\kappa_s^{3/2}\Gamma\left( \kappa_s -\frac{1}{2} \right)} \sum^\infty_{n=0} (-1)^n \left(\frac{\kappa_s}{\xi_s^2}\right)^{n+\frac{1}{2}} \times \notag \\
    &\left[ 
        \frac{\Gamma\left( n+\frac{1}{2} \right)}{\cos(\kappa_s\pi)\Gamma\left( n+\frac{1}{2}-\kappa_s \right)}
        +\tan(\kappa_s\pi)\frac{\Gamma(\kappa_s+n+1)}{\Gamma(n+1)} \left(\frac{\kappa_s}{\xi_s^2}\right)^{\kappa_s+\frac{1}{2}}
    \right].
    \label{eq:zs-ex}
\end{align}
where $|\arg(\xi_s^2/\kappa_s)|<\pi$.
Eq. \eqref{eq:zs-ex} coincides with the results in Refs. \onlinecite{Mace1995} and \onlinecite{Mace2009}.
We must emphasize that it is the expansion \eqref{eq:zs-ex} that diverges for half-integer $\kappa_s$. 
The hypergeometric representation of $Z(\kappa_\sigma,\xi_\sigma)$ \eqref{eq:Z-2f1}, as well as the definition \eqref{eq:Z-int}, converges for all kappa indices.

From Sec. \ref{sec:num-analy}, we find $\kappa_s$ should sufficiently approach $3/2$ to ensure the existence of the weakly damped EAWs.
When $\kappa_s \rightarrow 3/2$, $\tan(\kappa_s\pi)$ is a large factor.
Thus, the second term in the square bracket of Eq. \eqref{eq:zs-ex} cannot be omitted.
It leads to the expansion of Eq. \eqref{eq:pdr-re} must have the term including $(\kappa_s/\xi_s^2)^{\kappa_s+1/2}$.
As a result, the analytic EAW dispersion cannot be derived explicitly due to this term.

\bibliography{mylib}
\end{document}